\RequirePackage{fix-cm}
\documentclass[smallextended]{svjour3}       
\smartqed  
\usepackage{graphicx}
\begin{document}

\title{Reconsidering  the relation between  ``matter wave interference'' and ``wave-particle duality''
}

\titlerunning{Reconsidering wave-particle duality}        

\author{Lukas Mairhofer         \and  Oliver Passon
}


\institute{ L. Mairhofer \at
              University of Applied Sciences Technikum Wien\\
              Department Applied Mathematics and Physics\\
             Tel.: +43 6504545262\\
              \email{mairhofe@technikum-wien.at}  \\
              ORCID: 0000-0003-2935-8156
           \and
         O. Passon \at
              Bergische Universit\"at Wuppertal \\
	School for mathematics and natural sciences\\
              Tel.: +49 202 4392490\\
              \email{passon@uni-wuppertal.de} \\
              ORCID: 0000-0003-1198-5100
}

\date{Received: date / Accepted: date}

\maketitle
\begin{abstract}
Interference of more and more massive objects provides a spectacular confirmation of quantum theory. It is usually regarded as support for ``wave-particle duality" and in an extension of this duality even as support for ``complementarity". We first give an outline of the historical development of these notions. Already here it becomes evident that they are hard to define rigorously, i.e. have mainly a heuristic function. Then we discuss recent interference experiments of large and complex molecules which seem to support this heuristic function of ``duality". However, we show that in these experiments the diffraction of a {\em delocalized} center-of-mass wave function depends on the interaction of the {\em localized} structure of the molecule with the diffraction element. Thus, the molecules display ``dual features" at the same time, which contradicts the usual understanding of wave-particle duality. We conclude that the notion of ``wave-particle duality" deserves no place in modern quantum physics.

\end{abstract}
\medskip
\textbf{Declaration:}
Funding: no. Conflicts of interest/Competing interests: no. 
Availability of data and material: n/a. 
Code availability: n/a\\

\newpage 
\section{Introduction}
\label{intro}
In 2002 the journal  {\em Physics World} asked its readers to vote for the most beautiful physics experiment. A  majority choose the ``double-slit experiment with electrons" \cite{crease2002}. Famously, Richard Feynman called this experiment the ``only mystery" of quantum physics in his lectures \cite{feynman}. However, the Feynman lectures go on to state:
\begin{quote}
We should say right away that you should not try to set up this experiment. This experiment has never been done in just this way. [...] 
We are doing a ``thought experiment", which we have chosen because it is easy to think about. We know the results that would be obtained because there are many experiments that have been done, in which the scale and the proportions have been chosen to show the effects we shall describe.
\end{quote} 
Obviously, Feynman (in 1964 when preparing the third Volume of his lectures) was  unaware that this experiment had been successfully performed by Claus J\"onsson \cite{joensson,joensson1} already in 1959. 

Since then, more refined versions of these ``matter wave interference" experiments with electrons have been conducted \cite{merli,tonamura}. In addition the interference of more massive and extended  objects has been performed, e.g. using neutrons \cite{rauch1974,zeilinger}, atoms \cite{keith,carnal} or small \cite{schoellkopf} and large molecules \cite{arndt}.  As a preliminary end point of this development one could recently witness the demonstration that even a native polypeptide can be brought into interference \cite{pep}. 

It is curious to note that these most advanced experiments in quantum-interferometry and nanotechnology are still discussed in terms of ``complementarity" or ``wave-particle duality"  -- notions that have fallen from grace in other quarters of  quantum theory and are otherwise discussed in introductory textbooks only. Apparently these experiments are viewed as mere technical improvements of the double-slit experiment with electrons, i.e. no novel conclusions of fundamental and interpretational nature are drawn from them. We challenge this received view and scrutinize the alleged support that these experiments provide for those notions of early quantum theory. 

We  start in Sec.~\ref{roots} by giving a brief history of wave-particle duality  and the problems to provide  a rigorous foundation for this concept. However,   duality might still serve a heuristic function -- vaguely put as: ``quantum objects behave (dominantly) either as waves or as particles''.  Apparently, such a heuristic reading of  wave-particle duality is  also  supported by the  recent matter wave interference experiments. We will demonstrate in Sec.~\ref{exp} that  the interpretation of these experiments is more subtle and that   they do not support this heuristic reading of  wave-particle duality. In brief, we will show that these experiments apply an optical grating that requires a localized molecular structure for its functioning while the center-of-mass wave function is  delocalized at the same time. Thus, the heuristic notion that either particle- or wave-like aspects dominate breaks down.     In Sec.~\ref{sum} we summarize our results and argue that the notion of wave-particle duality should be discarded, since it implicitly assumes a space-time embedding which is not supported by quantum theory.  Furthermore the evaluation of recent matter wave experiments shows that both notion are required for their description at the same time rather than excluding each other. 



\section{Wave-particle duality and complementarity: Origin and some later developments\label{roots}}


The notion of a duality (or ``dualism'' -- we use these two terms interchangeably) of particles and waves first emerged in the study of {\em light}. 
While Einstein's light quantum hypothesis from 1905 already hints at a ``particle-like" aspect of light, historians of science usually trace the earliest reference of a ``true duality"  to Einstein's Salzburg lecture from 1909 \cite{einstein1909,jammer66,klein64}. 
There he calculated the mean square fluctuation of black-body radiation according to Planck's law. The result shows two terms.  One is proportional to the mean energy itself ($\propto \langle E\rangle$), i.e. showing a particle signature and could also be derived from the Wien law of black-body radiation. The  other term is proportional to the square of the mean energy ($\propto \langle E\rangle^2$), thus exhibiting a wave signature and follows when applying the Rayleigh-Jeans law. The presence of these two terms in the black-body spectrum as described by Planck's law thus  expresses the fact that technically Planck's law interpolates between the Wien and the Rayleigh–Jeans law (although this was not Planck's route for discovering the law). Einstein concludes in 1909 that ``the next stage in the development of theoretical physics will bring us a theory of light that can be understood as a kind of fusion of the wave and emission theories of light" \cite[p. 817]{einstein1909}. However, this quotation shows that  Einstein's interpretation of the result was not ``dualistic", as elaborated e.g. by Kojevnikov in \cite{kojevnikov}.

Famously, the acceptance of Einstein's light-quantum hypothesis was very slow and especially Bohr rejected it until 1925 \cite[p. 346]{jammer66}. Only after embracing the light quantum  he  developed his notion of ``complementarity", i.e. the need for descriptions that are mutually exclusive, but equally necessary.  As Jammer puts it \cite[p. 345]{jammer66}: ``Bohr's conception of  complementarity originated from his final acceptance of the wave-particle  duality."

When Bohr introduced the concept of complementarity in his Como lecture from 1927 \cite{bohr28} he used the mutual exclusiveness of space-time and causal descriptions as his first example but continued by discussing the particle- and wave-description of light {\em and} matter immediately after.  Further more, Bohr  repeatedly stressed the relation to mutually exclusive experimental arrangements for complementary properties.\footnote{We should note that, as Landsman puts it,  ``Bohr never gave a precise definition of `complementarity',  but  restricted himself to the analysis of a number of examples" \cite[p. 441]{landsman}. Also Purrington remarks that, for Bohr, complementarity functioned at multiple levels ``which makes it uncommonly difficult to state it succinctly"  \cite[p. 228]{purrington}. Rosenfeld -- a Bohr disciple whose opinions on Bohr are usually viewed as authoritative -- wrote: ``Complementarity is no system, no doctrine with readymade precepts. There is no via regia to it; no formal definition of it can even be found in Bohr’s writings, and this worries many people" \cite[p. 532]{rosenfeld}. We will come back to these  difficulties later.

}

That Bohr could discuss the complementarity not only for descriptions of light but also for matter is certainly due to the fact that Schr\"odinger had developed wave-mechanics already in 1926. However, the speculations about wave-like aspects of material objects had started already in 1923 with Louis de Broglie publishing three short articles in {\em Comptes Rendus} followed by a note in {\em Nature} \cite{broglie}.\footnote{As noted by Howard \cite{howard}, Einstein had toyed with the idea of a duality with respect to material particles around the same time.} This generalization of wave-particle duality famously introduced the de Broglie wave-length of a material particle having momentum $p$:
\begin{eqnarray}
\lambda=\frac{h}{p}. \label{dbr}
\end{eqnarray}
At that time the experimental support for these ``matter waves" was only indirect (see \cite{medicus} for an insightful discussion) and the final  confirmation had to wait until the work of Davisson and Germer in 1927 \cite{dg27} (preliminary announced in April 16, 1927). 

Gehrenbeck \cite{gehrenbeck78}  notes that the matter wave experiments of Davisson and Germer  gained immediate  acceptance. 
This was surely supported by the fact that this concept had successful applications already: In 1925 Einstein's (second) paper on the quantum theory of the monoatomic ideal gas \cite{einstein25} referred to de Broglie's thesis from 1924 when dealing with energy fluctuations of the  ideal gas (similar to the problem dealt with in 1909 with respect to radiation). This work was an important inspiration to   Erwin Schr\"odinger when developing wave-mechanics in 1926 \cite{jammer66}.

So, apparently there is a rather coherent development from Einstein (1905/09) over de Broglie (1923), Schr\"odinger (1926) and Davisson \& Germer (1927) to a full acceptance of the wave-particle duality for light {\em and} matter leading to  Bohr's notion  of  complementarity in 1927.

But there is a caveat with regard to Bohr's ``complementarity".  According to the widespread view the double-slit  experiment with electrons provides the paradigmatic  example for mutually exclusive  experimental arrangements which reveal complementary properties.  For example Michael Dickson puts it this way \cite[p. 345]{comp}:
\begin{quote}
[...] when we measure a wavelike property of particles (interference), we get wave-like behavior (interference pattern), while when we measure a particle-like property of particles (which slit a particle traverses), we get particle-like behavior (no interference pattern).  
\end{quote}
However, it has been argued repeatably (see e.g. Ballentine \cite[p. 4]{ballentine}, Born, Landau \& Lifshitz, Feynman or Heisenberg  \cite[footnote 43]{held}) that  the discrete detection of the electrons (assuming a beam of low intensity) provides evidence for a particle-like property within the double-slit experiment as well. In this sense the double-slit demonstrates  {\em both} properties in the very same experimental arrangement -- which  compromises  the above  reading of ``complementarity".  

This problem did not go unnoticed by Bohr himself who famously discussed the double-slit experiment   along similar  lines in his later writings. Presumably for this reason  Bohr did not use the ``wave-particle duality" as an example for ``complementarity" after 1935 \cite{held,pb}. Note, that this does not invalidate the concept of ``complementarity" as such  but only the kind of ``wave-particle complementarity" which unfortunately happens to be the most popular (not only) among textbook writers. Other versions (like the ``complementarity between causation and spatiotemporal location" or between incompatible variables) can be   maintained \cite{bitbol2016}. 

Likewise, one may rescue some sort of duality, given that  the wave- and particle aspects  play out at least at {\em different} stages of the same experiment (i.e. double-slit and detection screen). That is, the duality-slogan``quantum objects behave (dominantly) either as waves or as particles'' could apparently be maintained in this qualified sense.\footnote{In fact, there is an extensive literature  on the issue whether the trade-off between which-way information at the slit and the visibility of the interference fringes at the screen can be made quantitative. Jaegger, Shimony and Vaidman \cite{jsv} (and Englert independently \cite{englert}) could show that the quantities ``distinguishability'' ($\cal{D}$) and ``visibility'' ($\cal{V}$) obey the so-called ``wave-particle duality relation'' (WPDR): ${\cal{D}}^2+{\cal{V}}^2\le 1$. Meanwhile similar WPDRs have been formulated and originally they were viewed as conceptually independent from uncertainty relations. However, already D\"urr and Rempe found connections between certain WPDRs and Robertson-type uncertainty relation involving the standard deviation \cite{duerr}. These uncertainty relations are known to be just special cases of so-called entropic uncertainty relations (EUR) which apply entropy functions (e.g. the Shannon entropy) to quantify the uncertainty (see \cite{coles2017} for an overview on EUR and their applications). Applying EUR on probability distributions which encode which-way information or fringe visibility respectively, Coles et al. \cite{coles} could derive a whole class of WPDRs. This result indicates that wave-particle duality relations are fully equivalent to entropic uncertainty relations, i.e. there is no need to invoke an independent ``principle of complementarity'' or ``duality''  here.\label{fn:eur}} 



However, the status of the wave-particle duality of light changed significantly with the advent of quantum electrodynamics. In fact, only this theory provides the framework for discussing photons properly which were a foreign body in non-relativistic quantum mechanics anyway. For example Walter Heitler states already in 1936 with respect to the dual nature of light \cite[p. 63f]{heitler}: 
\begin{quote}
This analogy was extraordinarily fruitful in the   development of the quantum theory, but it should not now be overstressed. 
[...] there is no indication that, for instance, the idea of the `position of a light quantum' (or the `probability for the position') has any simple physical meaning. 
\end{quote} 
In the footnotes Heitler refers to a result by Landau and Peierls from 1930 which showed  that there is no wave function for the photon with probability interpretation in 3-space (see \cite[p. 10ff]{peierls} for a less technical version of the same argument). Newton and Wigner could show in 1949 that there is no position operator for the photon \cite{newton} which substantiates Heitler's claim further. In addition, most phenomena which have been historically attributed to the alleged ``particle-nature" of light (including e.g. the photoelectric effect and Compton scattering) can be explained semi-classically -- and this was realized already in 1926 \cite{milonni}. 

There are certainly effects like spontaneous emission or the Lamb shift which need a quantized radiation field for their explanation. The photon can then be identified with the state of the radiation field having the occupation number one. However, the only particle-like property of this QED photon is the discreteness of the spectrum of the number operator and the corresponding Fock-space representation.  

Further developments shed new light on the  wave-particle duality of matter. A prominent critic of  duality of matter was the late Alfred Land\'e. He  claimed in the 1970s that the de Broglie relation does not behave properly under Galilei transformations, since momentum and wavelength transform differently (i.e. $p^\prime =p+mv$ and $\lambda^\prime =\lambda$; here $v$ denotes the relative velocity between the primed and unprimed frame-of-reference). He concluded that therefor Equ.~\ref{dbr}  has to be dismissed as a foundation of quantum theory \cite{lande} which strongly challenges the basis for the wave-particle duality (of matter). 

This so-called Land\'e-paradox (or rather pseudo-paradox) was soon resolved by Jean-Marc L\'evi-Leblond \cite{ll}. He pointed out, that the assumption $\lambda^\prime =\lambda$ holds only for ``classical wave functions", while for the Galilei-transformation of the complex valued wave function of quantum mechanics, $\psi$,  an additional space-time dependent phase factor enters.\footnote{Given that all predictions of quantum mechanics are based on $|\psi|^2$, Galilei-invariance requires only the modulus of the wave function to remain unchanged.} For the transformed wavelength one finally arrives at \cite{ll,ll2}
\begin{eqnarray} 
\lambda^\prime = \frac{\lambda h}{h+\lambda m v}, \label{lambda}
\end{eqnarray}
and this expression renders the de Broglie relation Galilei-invariant.\footnote{This change in wave-length has been experimentally confirmed \cite{shull64}.} However,  while refuting Land\'e's argument,  L\'evi-Leblond made a perhaps even strong{\-}er case against ``wave-particle duality". Eventually he showed that for de Broglie-Schr\"odinger ``waves'' $\lambda^{\prime}\not= \lambda$ holds, i.e. a basic requirement of a wave-length when applying a Galiei-transformation is violated. This obviously calls into question whether one is dealing here with waves at all. Consequently,  also L\'evi-Leblond concluded that ``wave-particle duality" is a problematic  concept which should be abandoned \cite{ll3}.



While all these arguments compromise a rigorous version of ``wave-particle duality'', it may still serve as an  heuristics. Especially this function is exploited in the textbook tradition and according to Wheaton ``students today are taught wave-particle duality'' \cite{wheaton}.  Apparently this heuristic function is particularly reinforced  by the recent  matter wave interference experiments. In the following section we  discuss these experiments more closely and show that on a more detailed description also  here the notion that quantum objects behave either particle- or wave-like is not supported.


 
\section{Most recent experiments on matter wave interference\label{exp}}
The Long Baseline Universal Matterwave Interferometer (LUMI) of the Vienna Quantum Nanophysics group currently holds the world record for the largest, heaviest and most complex objects exhibiting quantum superposition. This superposition is demonstrated by recording the interference pattern of molecules with a mass exceeding 25.000 amu \cite{lumi}. A dilute molecular beam emanates from a thermal source, passes through the interferometer and then is detected in a mass spectrometer. 

The interferometer operates in the near-field, where the interference pattern reproduces the structure of the diffraction element at certain distances behind the mask. These distances are integer multiples of the Talbot length \cite{talbot} $L_T = 2d/\lambda_{dB}$, with $d$ the grating period and $\lambda_{dB}$ the de Broglie wavelength \cite{broglie} associated with the center-of-mass motion of the molecule: $\lambda_{dB}=h/mv$. 
A quantum field theoretical description of such complex structures as those molecules has yet to be formulated. The matter wave picture that follows the concept of de Broglie, however, allows  to describe the interferometer in terms of wave optics. The complex-valued density operator that encodes the dynamics of the matter wave can be translated into real-valued Wigner-functions. Wigner functions represent the evolution of a quantum state in phase space and provide a more intuitive picture of quantum mechanical processes than a Hilbert space formulation. This allows a direct comparison of quantum and classical predictions. This translation requires a one-to-one map from the density operator $\rho$ to a distribution function $W(x,p)$. This map is given by the following axiomatic relations:
\begin{equation}
 W(x,p)=1/2\pi\hbar\int_R dq e^{-ixq/\hbar}\langle p-q/2|\rho|p+q/2\rangle\end{equation} in terms of the density operator in the momentum representation and 
\begin{equation}
W(x,p)=1/2\pi\hbar\int_R ds e^{-ixs/\hbar}\langle x-s/2|\rho|x+s/2\rangle
\end{equation} in the position representation \cite{nimmrichter}.

While a variety of mechanisms for producing molecular beams exist, the LUMI relies on a Knudsen cell as standard source. From the furnace a dilute molecular beam emanates, such that the individual objects do no interact with one another. Typical velocities of the molecules range from 100--200 m/s, with a thermal Boltzmann-distribution from which a section with a velocity spread of 10--20\% is selected by inserting three slits defining a parabola of free fall. While the current upper mass limit is 25.000 amu, the Fullerene C$_{60}$ with a mass of 720 amu is the working horse molecule for calibration and alignment purposes. Thus the interferometer operates with de Broglie wavelength of 50 fm to 5 pm, which is up to five orders of magnitude smaller than the size of the molecules itself, which is of the order of nanometers. 

\begin{figure}[th]
   \includegraphics[width=0.95\textwidth]{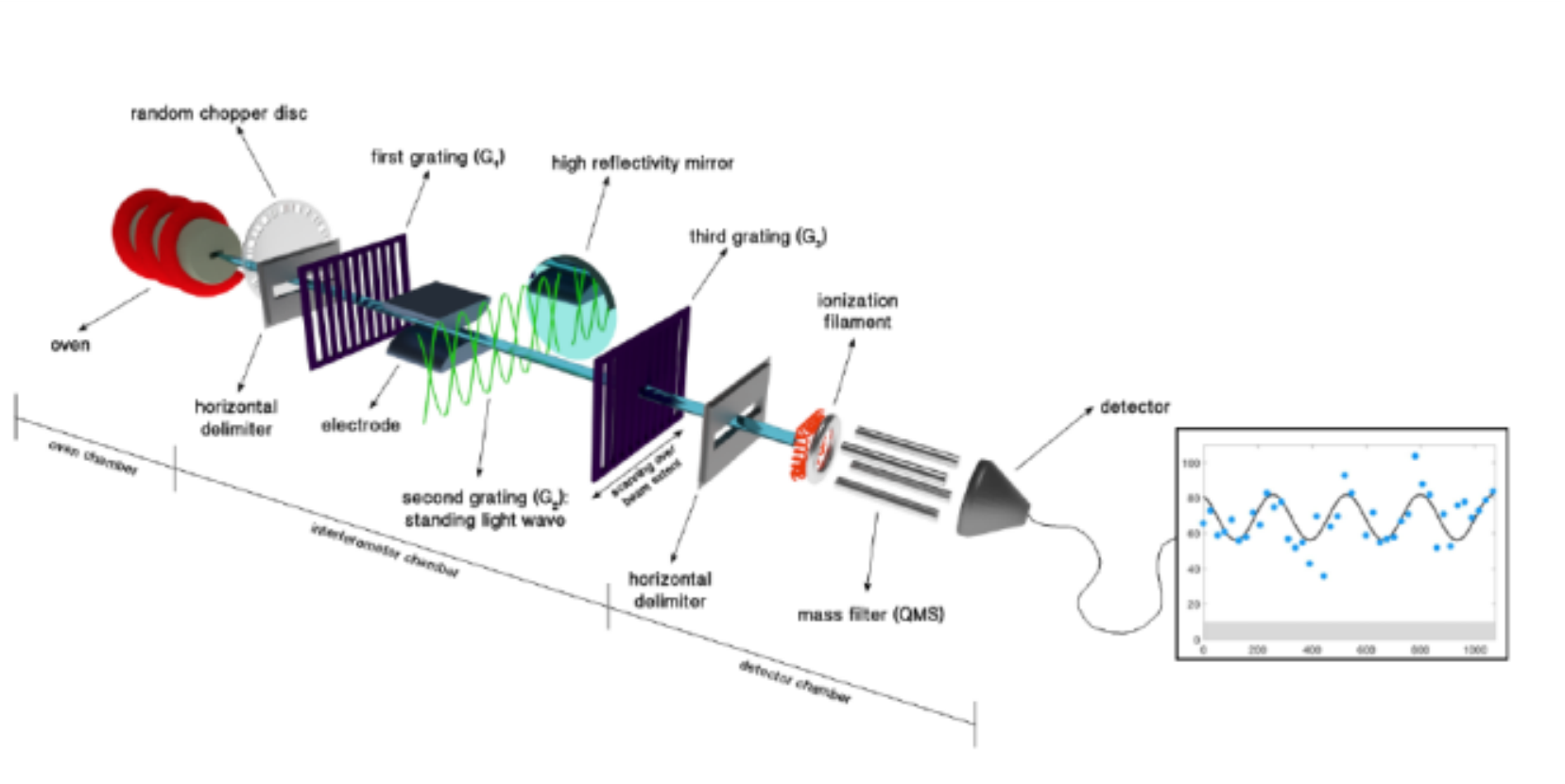}
 \caption{Kapitza-Dirac-Talbot-Lau interferometer (KDTLI): Molecular coherence is prepared by diffraction at each slit of the first grating G1. The coherence function spreads out and covers more than two anti-nodes of the optical grating G2. The spatially periodic phase imprinted by the standing light wave and subsequent interference lead to the formation of a molecular density pattern at the third grating, G3. It serves to mask the molecular fringe pattern, before the molecules are ionized and counted in the quadrupole mass spectrometer. Picture courtesy of Marion Romirer. 
 }
\label{fig:1}       
\end{figure}

The interferometer consists of three gratings that all have an equal period $d$ of 266 nm. The gratings are placed along the molecular beam at equal distances $L=1$ m to one another. Figure \ref{fig:1} shows a sketch of the experimental setup. The first and the third grating are material masks, photolithographically etched into a 10 nm thick layer of silicon nitride. 

The first grating acts as an array of point-like sources, imprinting transverse coherence onto the initially incoherent molecular beam. 
In terms of the Heisenberg uncertainty principle $\Delta x\Delta p\geq \hbar$ this spatial or transverse coherence $W_c$ results from the limitation of the molecule's transverse horizontal position $\Delta x$ to the extension of one of the slits of the grating, which are about 100nm wide. The resulting uncertainty of the molecules' momentum $\Delta p$ translates back into an uncertainty of the position of its center-of-mass, which grows linearly with distance $L$ behind the grating. 

In terms of wave optics transverse coherence depends on the size $s$ of the aperture of the wave source. The grating acts as an array of sources, each with a horizontal extension $s$ of 100nm.The resulting spatial coherence $W_c$ can be calculated from the Van Cittert-Zernike theorem \cite{colloqium}. This theorem states that a wavefront emanating from a spatially incoherent source will nevertheless exhibit spatial coherence when observed at a distance from the source that is large compared to the size of the source and the wavelength. This yields the same spatial coherence as predicted by Heisenberg's uncertainty: 
\begin{equation}W_c \,\propto\, 2\frac{L\lambda_{dB}}{s}.
\end{equation}
At the position of the second grating, the transverse coherence is at least two times larger than the period of the grating. In other words, transverse to its direction of motion the center-of-mass is delocalized over a thousand times the molecules size and up to $10^8$ times its de Broglie wavelength. 

The third grating acts as detection mask. A piezo scans the mask transverse over the molecular beam in steps of about 20 nm. Behind the interferometer the molecules are ionized by electron impact, mass selected in a quadrupole mass spectrometer and counted with a channeltron. The modulation of the count rate $S$ with the position of the third grating yields the interference pattern. 

The actual diffraction element is the central grating G2 consisting of a standing light wave, obtained by retro-reflecting a laser with a wavelength of 532 nm from a mirror. The intensity in the center of this standing wave is: $I=\frac{2P}{\Pi}w_xw_y$, with $P$ the laser power, $w_x$ the horizontal and $w_y$ the vertical waist. The laser interacts with the molecules via their optical polarizability $\alpha_{opt}(\omega)$ and their absorption cross section $\sigma(\omega)$. While for decades physicists diffracted light at matter, here large chunks of matter are diffracted at light. 

In LUMI, the interaction between the standing light wave and the molecule via its optical polarizability is the dominant diffraction mechanism. The rapidly oscillating electromagnetic potential of the standing wave induces an electric dipole moment in the passing molecule, which again couples to the external electromagnetic potential. The effect of this interaction can be described as a position dependent phase shift modulating the molecules' center-of-mass wave function. It depends on the intensity of the external field and the polarizability of the molecule, which encodes how easily and over which distance the external field shifts charges inside the molecule. The induced dipole moment $p$ depends on the number of charges $q$ as well as their separation $r$: $p=qr$. The separation induced by the external field depends on the spatial arrangement of the atomic cores, which determines the accessible electronic bands and the bond structure. 

This arrangement can only be described in terms of a {\em localized} molecule where the distances between the atom cores   are much smaller than the delocalization of its center-of-mass wave function. If the delocalization of the center-of-mass would alter the optical polarizability, molecules with different velocities and thus different de Broglie wavelengths would exhibit different interactions with the light grating. This is not observed. Thus the delocalized center-of-mass wave function covers several nodes of the standing light field, but the interaction with this field is described in terms of the molecules' localized structure. Here, two descriptions clash, the wave concept and the particle picture. While wave-particle duality claims that they are mutually exclusive, the description of the diffraction of complex molecules requires that we apply them both at the same time. 

Arguably, gratings made of light play an important role in atom interferometry as well \cite{rasel} and here, too, the interaction between the external field and the atom depends on its internal electronic structure. However, in the Heisenberg picture the atom's electronic structure can be described without reference to a localized system. As we shall show, this becomes implausible in the case of large and complex molecules, whose internal structure consists of an arrangement of atom cores and an electronic band structure that is strongly influenced by the spatial constellation of the cores. In addition to the 6 external degrees of freedom (translation and rotation), a structure consisting of $N$ atoms has $3N-6$ vibrational degrees of freedom. Since the molecules leave the thermal source with an internal temperature of several hundred Kelvin, all or at least the vast majority of those degrees of freedom are accessible. This has several important consequences for the diffraction of the center-of-mass of such an entity at a standing light wave. 

First of all, it excludes an interpretation of the interference pattern as resulting from the interference of multiple particles with one another. Interference of entities requires their indistinguishability, that is they have to be in the same state. 
For photons, electrons and atoms the demonstration of the interference of each object with itself requires careful preparation of a low-intensity beam. In the case of hot and complex molecules the chance that two molecules passing the grating at the same time are in the same state is much too small for accounting for the interference pattern, even for a beam of high intensity. 

The second point is, that the internal degrees of freedom of the system provide it with its own heat sink. Upon absorption of a photon, the molecule may redistribute the energy into its internal degrees of freedom and thus avoid re-emission of a photon. This leads to an increase of the particles internal temperature by exciting vibrational modes. Since the coherence length of the laser is orders of magnitude larger than the transverse extension of the light grating, the photon may be part of the incoming as well as of the reflected laser beam. Upon absorption, this superposition of momentum states is transferred to the molecule. It has been demonstrated that such a coherent absorption process acts as an additional beam splitting mechanism \cite{cotter}. This process, too, relies on a localized structure of atom cores. 

Finally, in contrast to atoms, the molecule's structure is dynamical itself. The distances between the atomic cores fluctuate on a sub-nanosecond timescale, depending on the internal temperature. These fluctuations constantly alter the accessible electronic bands and sometimes even the bond structure itself, influencing the electronic, magnetic and optical properties of the molecules. 

The molecule's configuration determines whether electronic states are localized to one atom core or delocalized over several atomic bonds. Take as an example the case of beta-Carotene. This molecule has several structural isomers -- it can form a straight line, called a trans-state, or one of the bonds can bend, which is called a cis-state. Here an isomerisation of the molecule from the trans- into one of its cis-states prevents the transfer of the delocalized pi-electrons across the whole molecule and therefore significantly reduces its response to the standing light-wave (\cite{vitamins}). 

For complex structures such as vitamins, simulations are required for an estimation of their optical, electronic and magnetic properties. Both semi-classical and quantum mechanical approaches depend on the assumption of localized atomic cores whose positions as well as dynamics determine the electronic band structure. Only when all of the six possible isomeric states of the molecule are taken into account, the simulation agrees with the experimental results (\cite{vitamins}). The dynamics of the localized structure of the molecule thus becomes visible in the interference pattern of its delocalized center-of-mass.


But not just the interaction of the delocalized center-of-mass wave function with the light grating depends on the molecule's internal localized structure. We can imprint a phase shift on the interference pattern evolving in free flight. This we achieve by applying a magnetic or electric deflection field. Since the famous experiments of Stern and Gerlach \cite{sterngerlach} deflectometry has developed into an important tool for determining the magnetic and electric properties of atoms, molecules and clusters \cite{dugourd,moro,schafer}. The fine-structured interference pattern provides a ruler on the nano-scale, thus significantly increasing the resolution of such deflection measurements. \cite{meterstick}.

In quantum assisted metrology the deflecting field is placed inside the interferometer, either between the first and the second, or between the second and the third grating \cite{electrode}, \cite{magnet}. Quantum assisted deflectometry has for example allowed to distinguish structural isomers \cite{isomers}, that is molecules consisting of the same atoms, but in a different arrangement. The effect of the magnetic or electric field on the evolution of the delocalized center-of-mass wave function again depends on the localized structure of the molecule. Although there is no doubt, that we can give quantum mechanical description of the electron's behaviour causing the electronic and magnetic properties, this description requires the localized structure of atomic cores \cite{vleck}. As a simple example, consider a diatomic molecule. Such a molecule exhibits axial symmetry instead of central symmetry. Here, the diamagnetic moment results not from the total orbital angular momentum of an electron, but only from the component which parallel to the axis of figure. The quantization of this component requires a well defined (i.e. localized) molecular structure, and thus a delocalization of the atomic cores would render the quantum mechanical description impossible. At the same time, the interaction of the external magnetic field with this localized internal structure of the molecule imprints a shift on the interference pattern of the delocalized center-of-mass that evolves in free flight \cite{magnetdeflectometry2}. Again, the description of the behavior of the wave function requires the concept of a particle-like structure.


\section{Summary\label{sum}}

The recent matter wave interference experiments  provide technological breakthroughs and a spectacular confirmation of quantum theory. However, as elaborated in Sec.~\ref{exp},  their interpretation  requires some care. With respect to the objective of our paper we emphasize: 
\begin{itemize}
\item The interference is detected with respect to only one degree of freedom, namely the center-of-mass motion. Only this allows to assign a simple de Broglie wavelength to the complex objects. 

\item   The action of the   optical grating can only be understood in terms of a localized interaction while the center-of-mass wave function is delocalized at the same time, hence there are  no clearly separated stages in which either the  ``particle-" or ``wave-like" behavior   dominates.  
\end{itemize}



We believe that it is time to reconsider the notion of wave-particle duality in principle. It is curious to discover the inconsistent views with regard to duality anyway:  to some it seems to be a manifestation of quantum weirdness still, some see it as a problem which has been solved and still others view it as a heuristic principle which has played out its role, i.e. neither an open or solved but no problem at all. John Hendry  sides with the last option when he writes \cite{hendry80}:
\begin{quote}
The wave-particle problem was never really resolved. [...] 
 But the duality problem had been absorbed into other issues. With their new quantum mechanics, Heisenberg and Pauli introduced a new conceptual framework in which a consistent structural description in classical space-time was no longer seen as necessary, and in which the wave-particle duality was no longer seen as problematic.
\end{quote}
To us, this quotation captures the key idea on how to think about ``wave-particle duality" today. To apply the notions of ``particle'' or ``wave''   implicitly assumes a space-time account of the process between preparation and measurement. However, quantum mechanics teaches us that such an account can not be given. While the Wigner-functions applied in the description of matter wave interference operate in space-time, they yield only a pseudo probability distribution as they can take negative values. 
While duality served an enormously useful purpose in the early development of quantum mechanics, it does not have a proper place in modern quantum theory (compare also Footnote~\ref{fn:eur}). 

Much of the  initial strength of wave-particle duality stemmed from the apparently unifying account with regard to light and matter, i.e. to parallelize photons with electrons and light waves with the de Broglie-Schr\"odinger matter waves. However, on closer scrutiny this  analogy breaks down in many respects (see Sec.~\ref{roots}). 
Where the notions of particle and wave apply in the description of state-of-the-art interference experiment with large and complex molecules, their duality turns out to be untenable (see Sec. 3).

Expressed pointedly, the continual charge against quantum theory of being ``weird" and ``bizarre" tells more about the plaintiff than the  defendant and  to discuss quantum theory against the backdrop of classical notions (like ``wave''  and ``particle'') shows simply a neglect of its autonomy.

\section*{Acknowledgment}
The idea of this paper was born at the conference dinner of the workshop ``How Quantum Mechanics changed Philosophy'' in Wuppertal early 2020. We thank  Marij van Strien for organizing this beautiful event.




\end{document}